\documentclass[twocolumn]{openjournal}
\usepackage[utf8]{inputenc}
\usepackage[english]{babel}
\usepackage{xcolor}
\usepackage{graphicx}
\usepackage{amssymb}
\usepackage{hyperref}
\usepackage{etoolbox}
\definecolor{linkcolor}{rgb}{0.0,0.3,0.5}
\hypersetup{unicode,colorlinks=true,linkcolor=linkcolor,citecolor=linkcolor,urlcolor=linkcolor}
\graphicspath{{./figs/}}
\newcommand{\nh}{N_{\mathrm H}}
\newcommand{\lbol}{L_{\mathrm{bol}}}

\providecommand{\arcsec}{^{\prime\prime}}
\providecommand{\arcmin}{^{\prime}}
\providecommand{\arcdeg}{^{\circ}}
\providecommand{\farcs}{.\!^{\prime\prime}}

\providecommand{\fs}{.\!^{\mathrm s}}
\begin{document}
\title{No Supersoft X-rays Before SN 2026ahvu: \\ Constraints on a Potentially Prompt Type Ia Progenitor}

\makeatletter
\clo@superscriptaddress
\renewcommand{\frontmatter@title@below}{\vspace*{6pt}}

\patchcmd{\doauthor}
  {\endgroup{\comma@space}{}\frontmatter@footnote}
  {\endgroup{}{}\frontmatter@footnote}
  {}{}
\makeatother

\author{\'{A}kos Bogd\'{a}n}
\email{akos@eurekasci.com}
\affiliation{Eureka Scientific, 2452 Delmer Street, Oakland, CA 94602}

\author{Tonima Tasnim Ananna}
\affiliation{Department of Physics \& Astronomy, Wayne State University, 42 W Warren Ave, Detroit, MI 48202, USA}

\begin{abstract}
The nearby Type~Ia supernova SN~2026ahvu offers a rare opportunity to probe its progenitor that may have evolved unusually rapidly. If the progenitor formed in either of the two young stellar clusters near the explosion site in NGC~1433, the delay from star formation to explosion could have been  $\sim$100~Myr or less. The distance to NGC~1433 remains uncertain, with published estimates of 9.1 and 18.9~Mpc. Our preliminary fit to the rising ATLAS light curve yields a distance of $17.7^{+1.5}_{-1.4}$~Mpc. In the classical single-degenerate scenario, a white dwarf steadily burning hydrogen accreted from a non-degenerate companion can emit detectable supersoft X-rays before explosion. 
We search for this emission using two archival \textit{Chandra} observations obtained before the explosion. The observations provide a combined exposure of 95.3~ks, with no $0.3-7$~keV photons detected within $2\arcsec$ of the supernova position.
For a 100~eV blackbody with Galactic foreground absorption, the 95\% bolometric luminosity limits are $6.3\times10^{37}$ and $2.7\times10^{38}\,\mathrm{erg\,s^{-1}}$ at 9.1 and 18.9~Mpc, respectively. At both distances, the observations exclude a hot, compact white dwarf undergoing steady hydrogen burning if the absorption is limited to the Galactic foreground. However, cooler, expanded, or more heavily absorbed photospheres remain viable. Together with the pre-explosion non-detections for SN~2011fe and SN~2014J, our results reinforce the absence of luminous supersoft progenitors in nearby SNe~Ia. This pattern is naturally consistent with double-degenerate explosions, although X-ray-faint single-degenerate configurations remain viable. Post-explosion X-ray observations can test for circumstellar interaction and help determine whether material around the progenitor could have obscured a luminous nuclear-burning white dwarf.
\end{abstract}

\maketitle

\section{Introduction}\label{sec:intro}

Modern transient surveys discover large numbers of supernovae each year. However, Type~Ia supernovae (SNe~Ia) within $\sim$$20$~Mpc with archival observations deep enough to constrain their progenitors remain extremely rare. Such observations offer a direct way to test models in which a white dwarf (WD) accretes from a non-degenerate companion \citep{whelan73} or explodes following the merger of two WDs \citep{iben84,webbink84}. A thermonuclear explosion can also occur below the Chandrasekhar mass if a helium-shell detonation triggers a second detonation in the carbon-oxygen core \citep{nomoto82,livne90}. For reviews of these and other progenitor pathways, we refer to \citet{maoz14,ruiter25}.

In the classical hydrogen-accreting single-degenerate scenario, a WD must retain enough of the transferred material to grow from its initial mass to near the Chandrasekhar limit, $M_{\rm Ch}\simeq1.4\,M_\odot$. Steady hydrogen burning provides an efficient route for this growth, since the accreted hydrogen is converted to helium without the repeated mass ejection associated with nova eruptions. The growth of the WD also depends on how much of the accumulated helium it retains \citep{nomoto07,wolf13}. At accretion rates of a few  $\times10^{-7}\,M_\odot\,\mathrm{yr^{-1}}$, the corresponding nuclear-burning luminosities are $\sim$$10^{37}-10^{38}\,\mathrm{erg\,s^{-1}}$. If the photosphere remains compact, its effective temperature can approach $10^6$~K, making the WD emit supersoft X-rays. Indeed, individual supersoft sources have been detected in nearby galaxies, including the Large Magellanic Cloud and M31 \citep{trumper91,distefano04}. Pre-explosion X-ray observations are therefore a key probe of the nuclear burning phase expected in the classical single-degenerate scenario.

If most SNe~Ia in early-type galaxies were produced by WDs that spent much of their mass growth as luminous supersoft sources, their combined soft X-ray emission would significantly exceed the total soft X-ray luminosity observed in those galaxies. Indeed, \citet{gilfanov10} found that the soft X-ray luminosities of nearby elliptical galaxies and bulges were a factor of $\sim$$30-50$ below the predictions for this scenario. They concluded that this classical channel could account for $<5\%$ of SNe~Ia in early-type galaxies. The comparison was subsequently extended to spiral and irregular galaxies \citep{bogdan11}. 

Several mechanisms have been proposed to reduce the combined X-ray luminosity anticipated for a population of nuclear burning WDs. Optically thick winds and recurrent nova eruptions can shorten the fraction of the accretion history spent in a luminous supersoft state \citep{hachisu10}. If the photosphere expands, much of the radiation can emerge in the extreme ultraviolet even at high bolometric luminosity \citep{nomoto07,lepo13}. Interstellar or circumstellar material can also absorb soft X-rays, although the column needed to obscure a source depends on the ionization state of the gas \citep{nielsen13,nielsen15}. Moreover, WD atmosphere models predict absorption edges and other spectral features that can alter the emergent soft X-ray flux \citep{suleimanov24}. However, shifting the emission to lower energies can still leave a substantial ionizing flux and produce observable nebular lines \citep{woods13,woods16}. Optical emission lines provide another way to constrain X-ray faint nuclear burning WDs. For example, \citet{johansson14} found that the weak He\,II $\lambda4686$ emission in passively evolving galaxies is inconsistent with a large population of steadily burning WDs over the temperature range they studied.

\begin{figure*}[!t]
\centering
\includegraphics[width=0.82\textwidth]{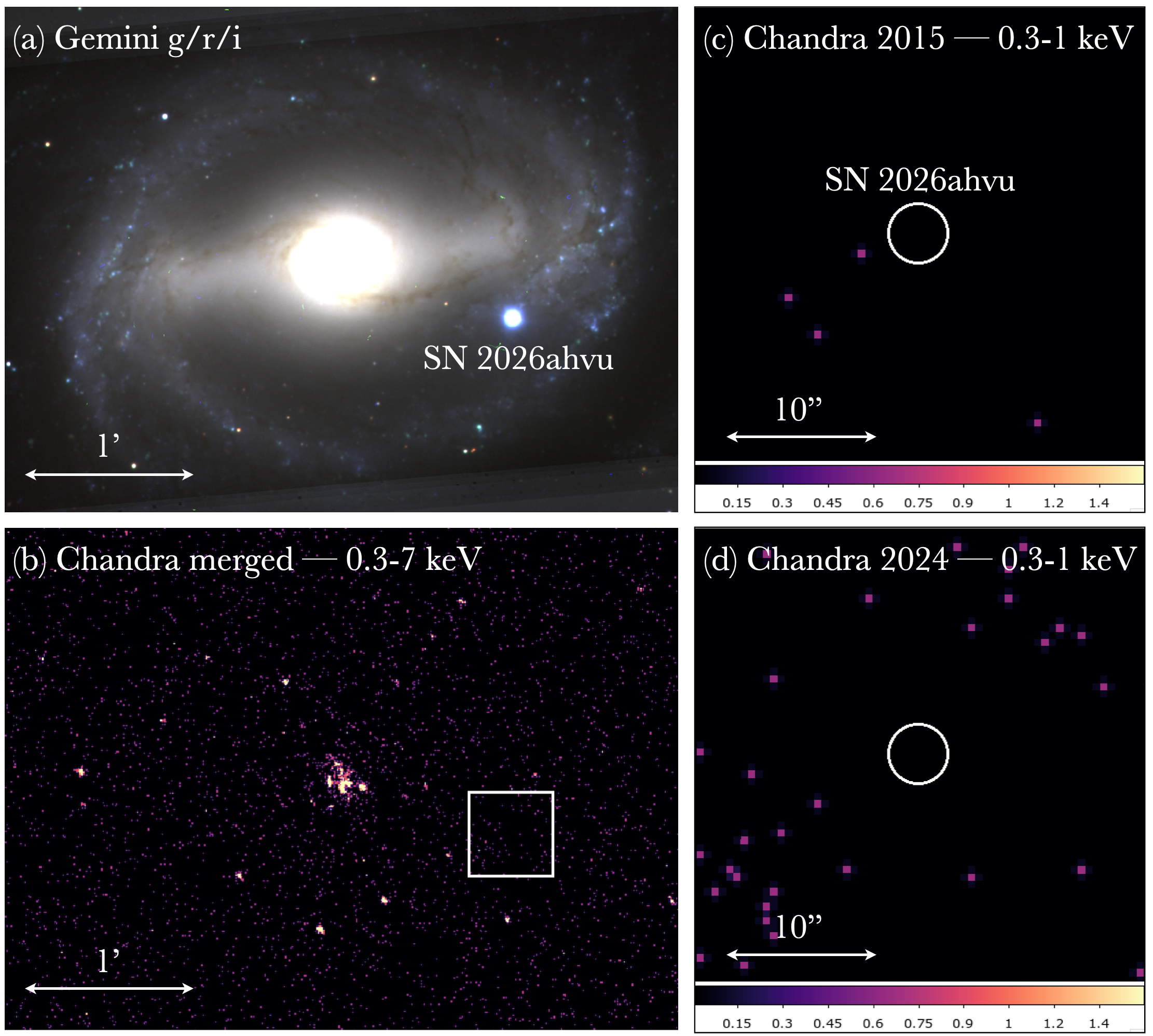}
\caption{Optical and X-ray images of SN~2026ahvu and its host galaxy, NGC~1433. (a) Gemini South/GMOS-S $gri$ color composite obtained on 2026 November~9, showing the supernova in the spiral disk. (b) Merged $0.3-7$~keV \textit{Chandra} image from the two archival observations. The white square marks the $30\arcsec\times30\arcsec$ region shown in the right panels. (c) and (d) $0.3-1$~keV \textit{Chandra} images from the 2015 (ObsID 16345) and 2024 (ObsID 28122) observations, respectively. The white circles mark the $2\arcsec$ radius source aperture centered on the SN position. Neither observation contains a photon within this aperture.}
\label{fig:field}
\end{figure*}

The progenitors of individual nearby SNe~Ia can be investigated by combining archival observations of the explosion site with observations obtained soon after the SN. Pre-explosion supersoft X-rays probe nuclear burning on the accreting WD, while post-explosion radio and X-ray observations probe the interaction of the ejecta with circumstellar material. \citet{nielsen12} searched archival \textit{Chandra} observations at the positions of ten SNe~Ia, including SN~2011fe in M101. For SN~2011fe, pre-explosion \textit{HST} images constrained the nature of the companion, and early radio and X-ray non-detections limited the density of circumstellar material \citep{li11,horesh12,chomiuk12}. For SN~2014J in M82, archival \textit{Chandra} observations excluded an unobscured, hot, compact steady burning WD, while cooler photospheres remained viable \citep{nielsen14}. The same observations revealed diffuse soft X-ray emission near the explosion site, possibly associated with recent star formation. 

SN~2026ahvu (ATLAS26ncp) offers an opportunity to test the progenitor of a potentially prompt SN~Ia with both pre- and post-explosion observations. ATLAS discovered the SN on 2026 November~2 in the barred spiral galaxy NGC~1433, and it was subsequently classified as a SN~Ia \citep{liu26,jacobson26,sarbadhicary26}. Its position is $\alpha=03^{\rm h}41^{\rm m}55\fs511$, $\delta=-47\arcdeg13\arcmin39\farcs39$ (J2000). Published distances to NGC~1433 include $9.1\pm1.0$~Mpc from the tip of the red giant branch \citep{sabbi18} and $18.94_{-0.56}^{+0.39}$~Mpc from the planetary nebula luminosity function \citep{scheuermann22}. We consider both distances throughout.

Using archival \textit{HST} and \textit{JWST} images, \citet{sarbadhicary26} identified two UV-bright clusters at projected angular separations of $0\farcs19$ and $0\farcs26$ from the SN. If the SN followed the spatial distribution of the older stellar population, the probability of finding it this close to a cluster would be  $\sim$0.4\%. If the progenitor formed in either cluster, the cluster age provides an estimate of the delay from star formation to explosion. Fits to the combined \textit{HST} and \textit{JWST} photometry give cluster ages of $37_{-5}^{+3}$ and $95_{-13}^{+15}$~Myr, respectively \citep{sarbadhicary26}. Both clusters show little inferred reddening and no associated H$\alpha$ emission. 

Multiwavelength follow-up of SN~2026ahvu has begun. \citet{cendes26} reported a $3\sigma$ flux-density upper limit of $<$0.22~mJy at 1.5~GHz about 5 days after discovery. Using the same two pre-explosion \textit{Chandra} observations analyzed here, \citet{kong26} reported a preliminary non-detection in the $0.3-7$~keV band and quoted luminosity limits for a power law and a 100~eV blackbody. For the blackbody, the reported $3\sigma$ upper limits on the 0.3--10~keV luminosity were $<1.2\times10^{39}$ and $<5.1\times10^{39}\,\mathrm{erg\,s^{-1}}$ at 9.1 and 18.9~Mpc, respectively.

Here we analyze the two archival \textit{Chandra} observations, obtained in 2015 and 2024, approximately 11.7~yr and 1.9~yr before discovery. The observations provide a total exposure of 95.3~ks. Neither observation contains a $0.3-7$~keV photon within a $2\arcsec$ source aperture. In this work, we account for the individual detector responses and derive 95\% upper limits on the bolometric luminosity as functions of photospheric temperature, absorbing column, and distance. We then compare these limits with predictions for compact and expanded nuclear-burning WDs to constrain the progenitor scenarios for SN~2026ahvu.

\begin{table}[!t]
\centering
\caption{Pre-explosion \textit{Chandra} observations of SN~2026ahvu.\label{tab:obs}}
\footnotesize
\setlength{\tabcolsep}{4pt}
\begin{tabular}{lccccc}
\hline\hline
ObsID & Date & CCD & Exp. & $N_{\rm src}$ & $N_{\rm bkg}$ \\
      &      &     & (ks) & & \\
\hline
16345 & 2015 Mar 04 & I3 & 48.6 & 0 & 7 \\
28122 & 2024 Dec 03 & S3 & 46.7 & 0 & 27 \\
\hline
\end{tabular}
\par\smallskip
\begin{minipage}{0.98\columnwidth}
\footnotesize
\textit{Note.} Counts were measured in the $0.3-2$~keV band. $N_{\rm src}$ is the number of counts within a $2\arcsec$ source aperture, while $N_{\rm bkg}$ is the number of counts in the  $5\arcsec-15\arcsec$ background annulus. 
\end{minipage}
\end{table}

\section{Chandra Observations and Data Analysis}
\label{sec:obs}

NGC~1433 was observed with \textit{Chandra} ACIS on 2015 March~4 (ObsID 16345) and 2024 December~3 (ObsID 28122). The observations provide a combined exposure of 95.3~ks (Table~\ref{tab:obs}). To analyze the data, we used CIAO~4.18 \citep{fruscione26} with CALDB~4.12.4.

We reprocessed both observations with the \texttt{chandra\_repro} task, applying the standard calibration and event filtering to obtain new level-2 event files. Neither observation showed significant background flaring in the $2.5-7$~keV light curve. We used \texttt{merge\_obs} to reproject the observations to a common tangent point and generate combined count images and exposure maps. The images were binned at the native ACIS pixel scale of $0\farcs492$. The main luminosity limits are derived in the $0.3-2$~keV band.

We characterized the \textit{Chandra} point spread function (PSF) at the SN position using the CIAO tools \texttt{psfsize\_srcs} and \texttt{src\_psffrac}. At 0.7~keV, the radii enclosing 90\% of the PSF are $1\farcs30$ and $1\farcs15$ for the 2015 and 2024 observations, respectively. We adopted a uniform $2\arcsec$ radius source aperture, enclosing approximately 96\% of the PSF in both observations. To measure the local background, we used an annulus centered on the SN with radii of $5\arcsec-15\arcsec$.

We generated auxiliary response files (ARFs) and redistribution matrix files (RMFs) separately for both observations. We calculated the expected source counts for each absorbed blackbody spectrum using the response and exposure of each observation. We modeled photoelectric absorption with \texttt{wabs} \citep{morrison83} and summed the expected counts from the two epochs to derive the bolometric luminosity limits.

For the Galactic foreground, we adopt $\nh=7.43\times10^{19}\,\mathrm{cm^{-2}}$  \citep{hi4pi16}. The low reddening inferred for the two nearby clusters suggests little dust extinction in the immediate vicinity of the SN \citep{sarbadhicary26}. However, the cluster photometry leaves the absorbing column along the progenitor sightline uncertain.  Therefore, we also consider a total column of $10^{21}\,\mathrm{cm^{-2}}$ to probe how additional absorption in the host galaxy or in circumstellar material would affect the luminosity limits.

\begin{figure*}[!t]
\centering
\includegraphics[width=0.75\textwidth]{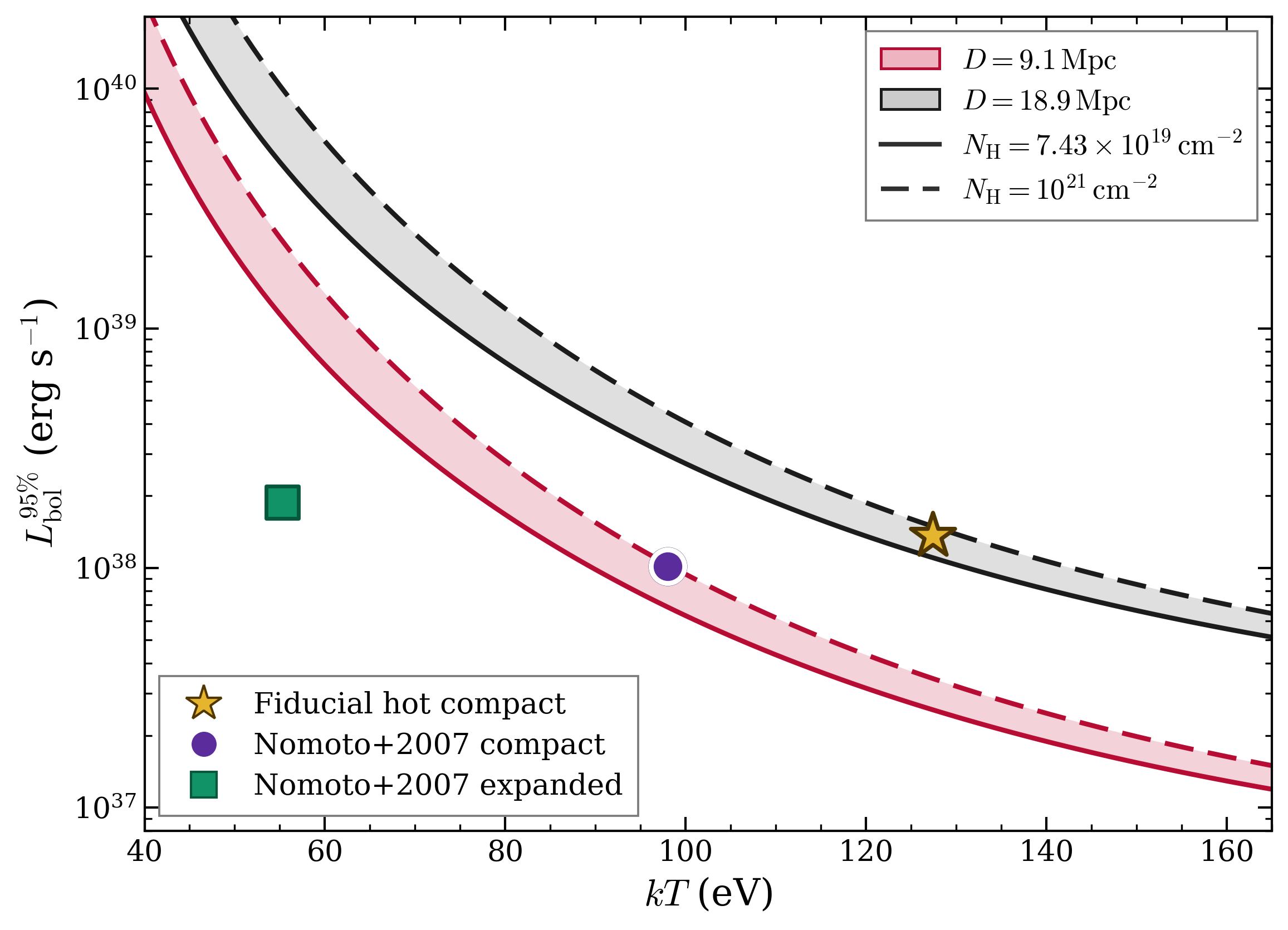}
\caption{95\% upper limits on the bolometric luminosity of a nuclear burning white dwarf as a function of blackbody temperature, derived from the $0.3-2$~keV \textit{Chandra} observations. The red and gray bands correspond to =9.1~Mpc and 18.9~Mpc distances, respectively. The lower boundaries assume a Galactic foreground, while the upper boundaries illustrate the effect of a higher assumed column. The symbols indicate our hot, compact near-Chandrasekhar benchmark and the compact and expanded steady burning models of \citet{nomoto07}.}
\label{fig:limits}
\end{figure*}

\section{Results}
\label{sec:results}

\subsection{X-ray non-detection and luminosity limits}

Figure~\ref{fig:field} shows the Gemini $gri$ color composite of NGC~1433, which locates SN~2026ahvu in the spiral disk, and the merged $0.3-7$~keV band \textit{Chandra} image of the same field. We note that the region surrounding the SN does not exhibit any bright X-ray point sources. Since most of the emission from a supersoft source is expected at low energies, we show separate $0.3-1$~keV images of the SN position for the 2015 and 2024 observations. Neither image contains a photon within the $2\arcsec$  source aperture. For the quantitative analysis, we adopt the broader $0.3-2$~keV band to include the high-energy tail of hotter nuclear-burning WD spectra. Neither observation contains a photon within the source aperture in this broader band.

With zero detected photons, we adopt a one-sided 95\% Poisson upper limit of 2.996 expected source counts. This corresponds to $0.3-2$~keV count-rate limits of $<$$6.17\times10^{-5}$ and $<$$6.41\times10^{-5}$~cts~s$^{-1}$ for the 2015 and 2024 observations, respectively. The combined exposure yields an upper limit of $<$$3.15\times10^{-5}$~counts~s$^{-1}$. To derive the bolometric luminosity limits, we fold each source spectrum through the individual instrumental responses and sum the predicted counts.

Figure~\ref{fig:limits} shows the 95\% upper limits on the intrinsic bolometric luminosity of a blackbody source as a function of temperature. We calculate these limits at 9.1~Mpc and 18.9~Mpc for two absorbing columns, namely utilizing the Galactic foreground value of $\nh=7.43\times10^{19}\,\mathrm{cm^{-2}}$ and a higher total column of $10^{21}\,\mathrm{cm^{-2}}$ (Section~\ref{sec:obs}). For Galactic foreground absorption, the bolometric luminosity limits at 9.1~Mpc are $<$$2.04\times10^{39}$, $<$$6.32\times10^{37}$, and $<$$2.56\times10^{37}\,\mathrm{erg\,s^{-1}}$ for blackbody temperatures of 50, 100, and 127~eV, respectively. At 18.9~Mpc, the corresponding limits are $<$$8.8\times10^{39}$, $<$$2.73\times10^{38}$, and $<$$1.10\times10^{38}\,\mathrm{erg\,s^{-1}}$. The limits rise steeply toward lower temperatures because progressively less of the bolometric luminosity emerges at energies where \textit{Chandra} remains sensitive.

Although the two observations have nearly equal exposures, we note that the 2015 observation provides most of the sensitivity to supersoft emission. By 2024, contamination on the ACIS optical blocking filters had substantially reduced the low-energy throughput \citep{plucinsky18,plucinsky22,plucinsky26}. For the 127~eV benchmark with Galactic foreground absorption, the 2015 observation contributes $\approx$85\% of the expected counts in the $0.3-2$~keV band.

\subsection{Constraints on nuclear burning white dwarfs}

We first consider a compact, near-Chandrasekhar-mass WD undergoing steady hydrogen burning. Motivated by the accretion rates in steady-burning envelope models \citep{nomoto07,wolf13}, we assume that hydrogen-rich material is processed at a rate of $\dot M_{\rm burn}=5\times10^{-7}\,M_\odot\,\mathrm{yr^{-1}}$ and has a hydrogen mass fraction of $X=0.72$. The resulting nuclear luminosity is $L_{\rm bol}\simeq X\epsilon_{\rm H}\dot M_{\rm burn}=1.36\times10^{38}\,\mathrm{erg\,s^{-1}}$, where $\epsilon_{\rm H}=6\times10^{18}\,\mathrm{erg\,g^{-1}}$ is the energy released per gram of hydrogen burned. For a photospheric radius of $R_{\rm ph}=2\times10^8$~cm, comparable to that of a $\sim$$1.38\,M_\odot$ WD \citep{nauenberg72}, we estimate an effective temperature of $kT_{\rm eff}\simeq127$~eV. With Galactic foreground absorption, this model predicts  $\approx$16 counts in the $0.3-2$~keV band at 9.1~Mpc and $\approx$3.7 counts at 18.9~Mpc (Table~\ref{tab:models}). Both exceed the 95\% Poisson upper limit.

\begin{table*}[!t]
\centering
\caption{Blackbody models for nuclear-burning white dwarfs and their predicted \textit{Chandra} counts.\label{tab:models}}
\footnotesize
\setlength{\tabcolsep}{6pt}
\begin{tabular}{lrrrrrrrr}
\hline\hline
Model & $kT_{\rm eff}$ & $R_{\rm ph}$ & $\lbol$ & $D$ & $N_{2015}$ & $N_{2024}$ & $N_{\rm tot}$ & $P_0$ \\
 & (eV) & ($10^8$ cm) & ($10^{38}$ erg s$^{-1}$) & (Mpc) & & & & \\
\hline
Compact benchmark & 127 & 2.0 & 1.36 & 9.1 & 13.5 & 2.5 & 16 & $1.2\times10^{-7}$ \\
 & & & & 18.9 & 3.1 & 0.57 & 3.7 & 0.025 \\
Nomoto compact & 98 & 2.9 & 1.01 & 9.1 & 4.0 & 0.44 & 4.4 & 0.012 \\
 & & & & 18.9 & 0.92 & 0.10 & 1.0 & 0.36 \\
Nomoto expanded & 55 & 12.5 & 1.87 & 9.1 & 0.49 & 0.01 & 0.50 & 0.61 \\
 & & & & 18.9 & 0.11 & $<0.01$ & 0.12 & 0.89 \\
\hline
\end{tabular}
\par\smallskip
\begin{minipage}{0.92\textwidth}
\footnotesize
\textit{Note.} Predicted counts assume persistent emission and Galactic foreground absorption in the $0.3-2$~keV band. 
\end{minipage}
\end{table*}

We next consider two steady hydrogen-burning envelopes calculated by \citet{nomoto07} for a $1.35\,M_\odot$ WD. At an accretion rate of $3.5\times10^{-7}\,M_\odot\,\mathrm{yr^{-1}}$, their model has a compact photosphere with $kT_{\rm eff}\simeq98$~eV and $L_{\rm bol}\simeq1.01\times10^{38}\,\mathrm{erg\,s^{-1}}$. At $6.0\times10^{-7}\,M_\odot\,\mathrm{yr^{-1}}$, the photosphere is more extended, with $kT_{\rm eff}\simeq55$~eV and $L_{\rm bol}\simeq1.87\times10^{38}\,\mathrm{erg\,s^{-1}}$. Representing each envelope by a blackbody at its effective temperature and luminosity, we predict $\approx$4.4 and $\approx$0.5 counts at 9.1~Mpc and $\approx$1.0 and $\approx$0.12 counts at 18.9~Mpc for the compact and expanded models, respectively. With Galactic foreground absorption, the compact envelope is excluded at 95\% confidence at 9.1~Mpc and remains consistent with the X-ray non-detection at 18.9~Mpc. Even at 9.1~Mpc, increasing the total column to $\nh=10^{21}\,\mathrm{cm^{-2}}$ lowers its predicted yield to 2.95 counts, just below the 95\% upper limit. Thus, the expanded envelope remains viable at both distances.

\section{Discussion}
\label{sec:discussion}

\subsection{Constraints on the progenitor photosphere}

Our results demonstrate that the X-ray detectability of a steadily burning WD depends strongly on its photospheric structure, rather than simply on its bolometric luminosity. An expanded photosphere can substantially suppress the supersoft emission even when the nuclear-burning luminosity remains high. Consequently, the pre-explosion \textit{Chandra} observations constrain compact and hot progenitors, but do not exclude steady nuclear burning in more extended or absorbed configurations.

The sensitivity of the 127~eV model to additional absorption differs markedly between the two distances. For the 127~eV model, the total column required to bring the predicted count yield down to the exclusion threshold is $\nh\simeq6.7\times10^{21}\,\mathrm{cm^{-2}}$ at 9.1~Mpc and $\nh\simeq7.1\times10^{20}\,\mathrm{cm^{-2}}$ at 18.9~Mpc. Photospheric expansion has a similarly strong effect. At a fixed luminosity of $1.36\times10^{38}\,\mathrm{erg\,s^{-1}}$ and with Galactic foreground absorption, the predicted yield reaches the exclusion threshold when the photospheric radius increases from $2.0\times10^8$~cm to about $4.6\times10^8$~cm at 9.1~Mpc or $2.3\times10^8$~cm at 18.9~Mpc. Thus, at the larger distance, a relatively small increase in photospheric radius is sufficient to hide an otherwise luminous nuclear-burning WD.

The role of absorption is also evident in earlier searches. The proximity and low foreground absorption of SN~2011fe yielded stringent limits on a supersoft progenitor \citep{nielsen12}. However, even the substantially deeper archival observations of SN~2014J were relatively insensitive to cooler photospheres because of the large absorbing column toward that event \citep{nielsen14}. For SN~2026ahvu, the Galactic foreground column is small, and the two nearby clusters show little reddening \citep{sarbadhicary26}. The absorbing column toward the progenitor remains uncertain, and ionized circumstellar gas may be less effective at absorbing supersoft X-rays than neutral gas \citep{nielsen13,nielsen15}.

The absence of pre-explosion supersoft emission is a natural outcome of the double-degenerate scenario. In this channel, two WDs can remain detached while their orbit shrinks through gravitational wave emission, without producing substantial accretion-powered X-ray emission \citep{iben84,webbink84,ruiter25}. In this respect, SN~2026ahvu resembles SN~2011fe and SN~2014J, for which archival \textit{Chandra} observations likewise did not reveal luminous supersoft progenitor \citep{nielsen12,nielsen14}. Together, these events establish a recurring observational pattern: despite differences in their environments and archival X-ray sensitivity, none shows evidence for the luminous, compact nuclear-burning WD expected in classical single-degenerate models. 

\subsection{The distance to NGC~1433}

The two published distances imply a factor of 4.3 difference in the inferred luminosities. The closer distance, $9.1\pm1.0$~Mpc, comes from a tip of the red giant branch measurement \citep{sabbi18}. The larger distance, $18.94_{-0.56}^{+0.39}$~Mpc, comes from the luminosity function of 90 planetary nebulae \citep{scheuermann22}. The optical light curve of SN~2026ahvu provides an independent, though preliminary, constraint on the distance to NGC~1433. We fitted the ten public ATLAS cyan- and orange-band measurements \citep{ahvuphot26} with the Pantheon+ SALT2 model and applied the Pantheon+/SH0ES luminosity calibration \citep{brout22,riess22}. The fit gives a distance modulus of $\mu=31.24\pm0.17$~mag, corresponding to $D=17.7^{+1.5}_{-1.4}$~Mpc, where the uncertainties are formal fit errors.  Of the published distances of NGC~1433, 18.9~Mpc provides a substantially better fit to the photometry than 9.1~Mpc. However, the available observations sample only the rise, and the inferred distance depends on how the model extrapolates to maximum light.

\subsection{Implications of a short delay time}

If SN~2026ahvu formed in one of the two nearby clusters, the cluster age constrains how rapidly its progenitor system had to evolve. The younger cluster would allow only a few tens of Myr to form a suitable WD and assemble a binary capable of exploding, while the older cluster would allow of order $10^8$~yr \citep{sarbadhicary26}. Once an accreting binary has formed, the final growth of the WD can be considerably faster. For example, a WD accreting at $5\times10^{-7}\,M_\odot\,\mathrm{yr^{-1}}$ would gain $0.2\,M_\odot$ in 0.4~Myr if it retained all the accreted material. Some helium-star donor models produce double detonations within $\sim$$10^8$~yr, while a double-WD system must form both WDs and undergo orbital decay within the available time \citep{ruiter11,ruiter25}. Thus, a physical association would constrain the time available to assemble the binary, while the X-ray data constrain whether a luminous hydrogen-burning photosphere was visible in the years before explosion.

Both clusters fall within the \textit{Chandra} source aperture. To estimate the scale of their possible X-ray contribution, we divide their combined fitted stellar mass, $3.3\times10^3\,M_\odot$ at 9.1~Mpc or $1.46\times10^4\,M_\odot$ at 18.9~Mpc, by an illustrative formation interval of 10~Myr \citep{sarbadhicary26}. Applying the high-mass X-ray binary scaling relation of \citet{mineo12} gives collective $0.5-8$~keV band luminosities of about $9\times10^{35}$ and $4\times10^{36}\,\mathrm{erg\,s^{-1}}$, respectively. For a power law with $\Gamma=1.7$ and Galactic foreground absorption, these luminosities correspond to  $\approx$0.3 expected counts in the $0.3-2$~keV band. As a caveat, we note that the relation was calibrated for star-forming galaxies, and the number of luminous binaries in two low-mass clusters depends on age and stochastic sampling. The calculation nevertheless shows that the archival observations would have little sensitivity to cluster emission at this level.

\subsection{Circumstellar interaction after the explosion}

The early MeerKAT non-detection constrains synchrotron emission from interaction between the SN ejecta and surrounding gas \citep{cendes26}. Converting the flux limit into a constraint on circumstellar density requires assumptions about the shock and the efficiency of synchrotron emission. A post-explosion \textit{Chandra} detection above the archival limits would establish new emission at the SN position. Its spectrum and evolution could help determine whether it arose from shocked gas or inverse-Compton scattering of SN photons by shock-accelerated electrons \citep{margutti14}. Conversely, a sufficiently deep non-detection would constrain the density over the radii reached by the shock, or the mass-loss rate for an assumed wind profile and velocity. In particular, the pre- and post-explosion measurements could test whether a wind dense enough to absorb supersoft X-rays before explosion would also produce detectable interaction afterward. An intrinsically cool photosphere could account for the archival non-detection even in the absence of such a wind.

\section{Conclusions}

We searched two archival \textit{Chandra} observations for X-ray emission from the progenitor of SN~2026ahvu, a SN~Ia near two young stellar clusters in NGC~1433. Neither observation contains a $0.3-7$~keV photon within $2\arcsec$ of the SN position. A preliminary fit to the rising ATLAS light curve gives an independent distance estimate of $17.7^{+1.5}_{-1.4}$~Mpc. For a 100~eV blackbody with Galactic foreground absorption, the 95\% bolometric luminosity limits derived in the $0.3-2$~keV band are $6.3\times10^{37}$ and $2.7\times10^{38}\,\mathrm{erg\,s^{-1}}$ at the two published galaxy distances of 9.1~Mpc and 18.9~Mpc, respectively.

With Galactic foreground absorption, the 127~eV compact WD model is excluded at both distances. The compact, steadily hydrogen-burning WD model of \citet{nomoto07}, with $kT_{\rm eff}\simeq98$~eV, is excluded at 9.1~Mpc for Galactic foreground absorption and remains consistent with the non-detection at 18.9~Mpc. Their cooler, expanded envelope remains viable at both distances. Thus, the observations constrain luminous, compact photospheres during the two archival epochs, while cooler or more absorbed nuclear-burning WDs remain viable. Together with SN~2011fe and SN~2014J, SN~2026ahvu reinforces the absence of luminous supersoft progenitors in pre-explosion X-ray observations of nearby SNe~Ia. This is consistent with double-degenerate explosions and increasingly challenges the simplest scenario in which Type~Ia supernovae are preceded by long-lived, unobscured nuclear burning on compact WDs. Nevertheless, the observations constrain the X-ray visibility of progenitors rather than uniquely identifying their evolutionary channel.

\section*{Acknowledgments}
This research has made use of data obtained from the Chandra Data Archive provided by the Chandra X-ray Center (CXC). The X-ray data were analyzed using the Chandra Interactive Analysis of Observations (CIAO) software. This work is based in part on observations obtained at the international Gemini Observatory (program GS-2026B-DD-104), a program of NSF's NOIRLab, which is managed by the Association of Universities for Research in Astronomy (AURA) under a cooperative agreement with the National Science Foundation on behalf of the Gemini Observatory partnership. \\

\section*{Data availability}
The observations are publicly available through the Chandra Data Archive as
\href{https://doi.org/10.25574/16345}{ObsID 16345} and
\href{https://doi.org/10.25574/28122}{ObsID 28122}.

\bibliographystyle{aasjournal}
\bibliography{references}

@misc{cendes26,
  author = {Cendes, Y. and Alexander, K. D. and Laskar, T. and Drout, M. and Goodwin, A.},
  year = {2026},
  title = {Radio non-detection of nearby {Type Ia SN 2026ahvu}},
  howpublished = {The Astronomer's Telegram, No. 18151},
  url = {https://www.astronomerstelegram.org/?read=18151}
}

@article{fruscione26,
  author = {Fruscione, A. and McDowell, J. C. and Burke, D. J. and others},
  year = {2026}, journal = {ApJ}, volume = {1005}, pages = {116},
  doi = {10.3847/1538-4357/ae6db4},
  title = {{CIAO} analysis software}
}

@article{gilfanov10,
  author = {Gilfanov, M. and Bogd{\'{a}}n, {\'{A}}.},
  year = {2010}, journal = {Nature}, volume = {463}, pages = {924},
  doi = {10.1038/nature08685},
  title = {An upper limit on the contribution of accreting white dwarfs to the type Ia supernova rate}
}

@misc{jacobson26,
  author = {Jacobson-Gal{\'{a}}n, W. and Li, M. and Hinds, K.-R.},
  year = {2026}, howpublished = {TNS AstroNote 2026-311},
  url = {https://www.wis-tns.org/astronotes/astronote/2026-311},
  title = {Classification of {SN 2026ahvu}}
}

@misc{kong26,
  author = {Kong, A. K. H.}, year = {2026},
  title = {X-ray upper limit on the progenitor of the {Type Ia} supernova {2026ahvu} in {NGC 1433}},
  howpublished = {The Astronomer's Telegram, No. 18148},
  url = {https://www.astronomerstelegram.org/?read=18148}
}

@misc{liu26,
  author = {Liu, C. and Smartt, S. J. and Smith, K. W. and others},
  year = {2026}, title = {{SN 2026ahvu} discovery},
  howpublished = {TNS AstroNote 2026-310},
  url = {https://www.wis-tns.org/astronotes/astronote/2026-310}
}

@article{morrison83,
  author = {Morrison, R. and McCammon, D.},
  title = {Interstellar photoelectric absorption cross sections, 0.03--10 keV},
  journal = {ApJ}, volume = {270}, pages = {119}, year = {1983},
  doi = {10.1086/161102}
}

@article{nauenberg72,
  author = {Nauenberg, M.}, year = {1972}, journal = {ApJ}, volume = {175}, pages = {417},
  doi = {10.1086/151568},
  title = {Analytic approximations to the mass-radius relation and energy of zero-temperature stars}
}

@article{nielsen12,
  author = {Nielsen, M. T. B. and Voss, R. and Nelemans, G.},
  year = {2012}, journal = {MNRAS}, volume = {426}, pages = {2668},
  doi = {10.1111/j.1365-2966.2012.21596.x},
  title = {Upper limits on bolometric luminosities of 10 Type Ia supernova progenitors from Chandra observations}
}

@article{nielsen14,
  author = {Nielsen, M. T. B. and Gilfanov, M. and Bogd{\'{a}}n, {\'{A}}. and Woods, T. E. and Nelemans, G.},
  year = {2014}, journal = {MNRAS}, volume = {442}, pages = {3400},
  doi = {10.1093/mnras/stu913},
  title = {Upper limits on the luminosity of the progenitor of Type Ia supernova SN 2014J}
}

@article{nomoto07,
  author = {Nomoto, K. and Saio, H. and Kato, M. and Hachisu, I.},
  year = {2007}, journal = {ApJ}, volume = {663}, pages = {1269},
  doi = {10.1086/518465},
  title = {Thermal Stability of White Dwarfs Accreting Hydrogen-rich Matter and Progenitors of Type Ia Supernovae}
}

@article{suleimanov24,
  author = {Suleimanov, V. F. and Tavleev, A. S. and Doroshenko, V. and Werner, K.},
  year = {2024}, journal = {A\&A}, volume = {688}, pages = {A39},
  doi = {10.1051/0004-6361/202449370},
  title = {Application of hydrostatic local thermodynamic equilibrium atmosphere models to interpretations of supersoft X-ray source spectra}
}

@article{wolf13,
  author = {Wolf, W. M. and Bildsten, L. and Brooks, J. and Paxton, B.},
  year = {2013}, journal = {ApJ}, volume = {777}, pages = {136},
  doi = {10.1088/0004-637X/777/2/136},
  title = {Hydrogen Burning on Accreting White Dwarfs: Stability, Recurrent Novae, and the Post-nova Supersoft Phase}
}

@article{sabbi18,
  author = {Sabbi, E. and Calzetti, D. and Ubeda, L. and others},
  year = {2018}, journal = {ApJS}, volume = {235}, pages = {23},
  doi = {10.3847/1538-4365/aaa8e5},
  title = {The Legacy ExtraGalactic UV Survey}
}

@misc{sarbadhicary26,
  author = {Sarbadhicary, S. K. and Henny, K. and Leroy, A. and others},
  year = {2026}, title = {{SN 2026ahvu} in {NGC 1433} May Be a Prompt {Type Ia} Supernova with a Delay Time Under 100 {Myr}},
  eprint = {2611.09875}, archivePrefix = {arXiv},
  url = {https://arxiv.org/abs/2611.09875}
}

@article{scheuermann22,
  author = {Scheuermann, F. and Kreckel, K. and Anand, G. S. and others},
  year = {2022}, journal = {MNRAS}, volume = {511}, pages = {6087},
  doi = {10.1093/mnras/stac110},
  title = {PHANGS--MUSE distances from the planetary nebula luminosity function}
}

@article{maoz14,
  author = {Maoz, D. and Mannucci, F. and Nelemans, G.},
  title = {Observational Clues to the Progenitors of Type Ia Supernovae},
  journal = {ARA\&A}, volume = {52}, pages = {107--170}, year = {2014},
  doi = {10.1146/annurev-astro-082812-141031}
}

@article{ruiter25,
  author = {Ruiter, A. J. and Seitenzahl, I. R.},
  title = {Type Ia supernova progenitors: a contemporary view of a long-standing puzzle},
  journal = {A\&ARv}, volume = {33}, pages = {1}, year = {2025},
  doi = {10.1007/s00159-024-00158-9}
}

@article{bogdan11,
  author = {Bogd{\'{a}}n, {\'{A}}. and Gilfanov, M.},
  title = {Soft-band X/K luminosity ratios in late-type galaxies and constraints on the population of supersoft X-ray sources},
  journal = {MNRAS}, volume = {412}, pages = {401--410}, year = {2011},
  doi = {10.1111/j.1365-2966.2010.17916.x}
}

@article{hachisu10,
  author = {Hachisu, I. and Kato, M. and Nomoto, K.},
  title = {Supersoft X-Ray Phase of Single Degenerate Type Ia Supernova Progenitors in Early-Type Galaxies},
  journal = {ApJL}, volume = {724}, pages = {L212--L216}, year = {2010},
  doi = {10.1088/2041-8205/724/2/L212}
}

@misc{lepo13,
  author = {Lepo, K. and van Kerkwijk, M. H.},
  title = {Rapidly accreting white dwarfs as supernova type Ia progenitors},
  year = {2013}, eprint = {1301.1368}, archivePrefix = {arXiv},
  url = {https://arxiv.org/abs/1301.1368}
}

@article{nielsen13,
  author = {Nielsen, M. T. B. and Dominik, C. and Nelemans, G. and Voss, R.},
  title = {Obscuration of supersoft X-ray sources by circumbinary material: A way to hide Type Ia supernova progenitors?},
  journal = {A\&A}, volume = {549}, pages = {A32}, year = {2013},
  doi = {10.1051/0004-6361/201219195}
}

@article{nielsen15,
  author = {Nielsen, M. T. B. and Gilfanov, M.},
  title = {Attenuation of supersoft X-ray sources by circumstellar material},
  journal = {MNRAS}, volume = {453}, pages = {2927--2936}, year = {2015},
  doi = {10.1093/mnras/stv1616}
}

@article{woods13,
  author = {Woods, T. E. and Gilfanov, M.},
  title = {He II recombination lines as a test of the nature of SN Ia progenitors in elliptical galaxies},
  journal = {MNRAS}, volume = {432}, pages = {1640--1650}, year = {2013},
  doi = {10.1093/mnras/stt586}
}

@article{woods16,
  author = {Woods, T. E. and Gilfanov, M.},
  title = {Where are all of the nebulae ionized by supersoft X-ray sources?},
  journal = {MNRAS}, volume = {455}, pages = {1770--1781}, year = {2016},
  doi = {10.1093/mnras/stv2423}
}

@article{li11,
  author = {Li, W. and Bloom, J. S. and Podsiadlowski, P. and others},
  title = {Exclusion of a luminous red giant as a companion star to the progenitor of supernova SN 2011fe},
  journal = {Nature}, volume = {480}, pages = {348--350}, year = {2011},
  doi = {10.1038/nature10646}
}

@article{horesh12,
  author = {Horesh, A. and Kulkarni, S. R. and Fox, D. B. and others},
  title = {Early Radio and X-Ray Observations of the Youngest Nearby Type Ia Supernova PTF 11kly (SN 2011fe)},
  journal = {ApJ}, volume = {746}, pages = {21}, year = {2012},
  doi = {10.1088/0004-637X/746/1/21}
}

@article{chomiuk12,
  author = {Chomiuk, L. and Soderberg, A. M. and Moe, M. and others},
  title = {EVLA Observations Constrain the Environment and Progenitor System of Type Ia Supernova 2011fe},
  journal = {ApJ}, volume = {750}, pages = {164}, year = {2012},
  doi = {10.1088/0004-637X/750/2/164}
}

@inproceedings{plucinsky18,
  author       = {Plucinsky, Paul P. and Bogdan, Akos and
                  Marshall, Herman L. and Tice, Neil W.},
  title        = {The Complicated Evolution of the {ACIS} Contamination
                  Layer over the Mission Life of the {Chandra}
                  X-ray Observatory},
  booktitle    = {Space Telescopes and Instrumentation 2018:
                  Ultraviolet to Gamma Ray},
  series       = {Proceedings of SPIE},
  volume       = {10699},
  pages        = {106996B},
  year         = {2018},
  doi          = {10.1117/12.2312748},
  eprint       = {1809.02225},
  archivePrefix = {arXiv}
}

@inproceedings{plucinsky22,
  author       = {Plucinsky, Paul P. and Bogdan, Akos and
                  Marshall, Herman L.},
  title        = {The Evolution of the {ACIS} Contamination Layer
                  on the {Chandra} X-ray Observatory through 2022},
  booktitle    = {Space Telescopes and Instrumentation 2022:
                  Ultraviolet to Gamma Ray},
  series       = {Proceedings of SPIE},
  volume       = {12181},
  pages        = {121816X},
  year         = {2022}
}

@misc{plucinsky26,
  author       = {Plucinsky, Paul P. and Ratzlaff, Peter W. and
                  Bogdan, Akos and Marshall, Herman L.},
  title        = {The Evolution of the {ACIS} Contamination Layer
                  on the {Chandra} X-ray Observatory from 2010 to 2026},
  year         = {2026},
  eprint       = {2608.14363},
  archivePrefix = {arXiv},
  primaryClass = {astro-ph.HE},
  note         = {SPIE Astronomical Telescopes and Instrumentation 2026,
                  Paper 14146-298}
}

@article{mineo12,
  author = {Mineo, S. and Gilfanov, M. and Sunyaev, R.},
  title = {X-ray emission from star-forming galaxies -- I. High-mass X-ray binaries},
  journal = {MNRAS},
  year = {2012},
  volume = {419},
  pages = {2095--2115},
  doi = {10.1111/j.1365-2966.2011.19862.x}
}

@article{margutti14,
  author = {Margutti, R. and others},
  title = {No X-Rays from the Very Nearby Type Ia SN 2014J: Constraints on Its Environment},
  journal = {ApJ},
  year = {2014},
  volume = {790},
  pages = {52},
  doi = {10.1088/0004-637X/790/1/52}
}

@article{hi4pi16,
  author = {{HI4PI Collaboration}},
  title = {{HI4PI}: a full-sky {H I} survey based on {EBHIS} and {GASS}},
  journal = {A\&A}, year = {2016}, volume = {594}, pages = {A116},
  doi = {10.1051/0004-6361/201629178}
}

@article{ruiter11,
  author = {Ruiter, A. J. and Belczynski, K. and Sim, S. A. and others},
  title = {Delay times and rates for Type Ia supernovae and thermonuclear explosions from double-detonation sub-Chandrasekhar mass models},
  journal = {MNRAS}, year = {2011}, volume = {417}, pages = {408--419},
  doi = {10.1111/j.1365-2966.2011.19276.x}
}

@article{whelan73,
  author = {Whelan, J. and Iben, Jr., I.},
  title = {Binaries and Supernovae of Type I},
  journal = {ApJ},
  year = {1973},
  volume = {186},
  pages = {1007--1014},
  doi = {10.1086/152565}
}

@article{iben84,
  author = {Iben, Jr., I. and Tutukov, A. V.},
  journal = {ApJS},
  year = {1984},
  volume = {54},
  pages = {335--372},
  adsurl = {https://ui.adsabs.harvard.edu/abs/1984ApJS...54..335I}
}

@article{webbink84,
  author = {Webbink, R. F.},
  title = {Double White Dwarfs as Progenitors of {R Coronae Borealis} Stars and Type I Supernovae},
  journal = {ApJ},
  year = {1984},
  volume = {277},
  pages = {355--360},
  adsurl = {https://ui.adsabs.harvard.edu/abs/1984ApJ...277..355W}
}

@article{nomoto82,
  author = {Nomoto, K.},
  title = {Accreting White Dwarf Models for Type I Supernovae. II. Off-Center Detonation Supernovae},
  journal = {ApJ},
  year = {1982},
  volume = {257},
  pages = {780--792},
  adsurl = {https://ui.adsabs.harvard.edu/abs/1982ApJ...257..780N}
}

@article{livne90,
  author = {Livne, E.},
  journal = {ApJL},
  year = {1990},
  volume = {354},
  pages = {L53--L55},
  adsurl = {https://ui.adsabs.harvard.edu/abs/1990ApJ...354L..53L}
}

@misc{ahvuphot26,
  author = {{Transients.fyi}},
  year = {2026},
  title = {{SN 2026ahvu} public photometry compilation},
  howpublished = {ATLAS cyan-band observation dated 2026 November 12, accessed 2026 November 13},
  note = {\href{https://transients.fyi/objects/tns/2026ahvu}{Public photometry page}},
  url = {https://transients.fyi/objects/tns/2026ahvu}
}

@article{riess22,
  author = {Riess, A. G. and Yuan, W. and Macri, L. M. and others},
  year = {2022}, journal = {ApJL}, volume = {934}, pages = {L7},
  title = {A Comprehensive Measurement of the Local Value of the Hubble Constant with 1 km s$^{-1}$ Mpc$^{-1}$ Uncertainty from the Hubble Space Telescope and the SH0ES Team},
  doi = {10.3847/2041-8213/ac5c5b}
}

@article{trumper91,
  author = {Tr{\"u}mper, J. and Hasinger, G. and Aschenbach, B. and others},
  title = {X-ray survey of the Large Magellanic Cloud by {ROSAT}},
  journal = {Nature},
  year = {1991},
  volume = {349},
  pages = {579--583},
  doi = {10.1038/349579a0}
}

@article{distefano04,
  author = {Di Stefano, R. and Kong, A. K. H. and Greiner, J. and others},
  title = {Supersoft X-Ray Sources in {M31}. {I}. A {Chandra} Survey and an Extension to Quasi-soft Sources},
  journal = {ApJ},
  year = {2004},
  volume = {610},
  pages = {247--260},
  adsurl = {https://ui.adsabs.harvard.edu/abs/2004ApJ...610..247D}
}

@article{johansson14,
  author = {Johansson, J. and Woods, T. E. and Gilfanov, M. and Sarzi, M. and Chen, Y.-M. and Oh, K.},
  title = {Diffuse Gas in Galaxies Sheds New Light on the Origin of Type {Ia} Supernovae},
  journal = {MNRAS},
  year = {2014},
  volume = {442},
  pages = {1079--1089},
  doi = {10.1093/mnras/stu907}
}

@article{brout22,
  author = {Brout, D. and Taylor, G. and Scolnic, D. and others},
  title = {The {Pantheon+} Analysis: SuperCal-fragilistic Cross Calibration,
           Retrained {SALT2} Light-curve Model, and Calibration Systematic Uncertainty},
  journal = {ApJ},
  year = {2022},
  volume = {938},
  pages = {111},
  doi = {10.3847/1538-4357/ac8bcc}
}
\end{document}